\documentclass[12pt]{article}
\usepackage{amsmath,amsfonts,amssymb}  

\makeatletter
\def\numberbysection{\@addtoreset{equation}{section}
    \def\theequation{\thesection.\arabic{equation}}}
\makeatother

\numberbysection

\newcommand{\be}{\begin{eqnarray}}
\newcommand{\ee}{\end{eqnarray}}
\newcommand{\non}{\nonumber}
\newcommand{\id}{\mathbb{I}}
\newcommand{\tr}{\mathop{\rm tr}\nolimits}
\newcommand{\str}{\mathop{\rm str}\nolimits}
\newcommand{\sgn}{\mathop{\rm sign}\nolimits}
\newcommand{\diag}{\mathop{\rm diag}\nolimits}

\begin{document}

\begin{titlepage}
\strut\hfill
\vspace{.5in}
\begin{center}

\LARGE A note on open-chain transfer matrices\\ 
\LARGE from {\it q}-deformed $su(2|2)$ $S$-matrices\\[1.0in]
\large Rajan Murgan\footnote{email: rmurgan@gustavus.edu}\\[0.8in]
\large Physics Department,\\ 
\large Gustavus Adolphus College,\\ 
\large 800 West College Avenue, St. Peter, MN 56082 USA\\
      
\end{center}

\vspace{.5in}

\begin{abstract}
In this note, we perform Sklyanin's construction of commuting open-chain/boundary transfer
matrices to the $q$-deformed $SU(2|2)$ bulk $S$-matrix of Beisert and Koroteev and a corresponding boundary $S$-matrix. 
This also includes a corresponding commuting transfer matrix using the graded version of the $q$-deformed bulk $S$-matrix. 
Utilizing the crossing property for the bulk $S$-matrix, we argue that the transfer matrix for both graded and non-graded versions 
contains a crucial factor which is essential for commutativity.
\end{abstract}

\end{titlepage}

\setcounter{footnote}{0}

\section{Introduction}\label{sec:intro}

Centrally extended $su(2|2)$ algebra (two copies) \cite{Be, AFPZ} is a key symmetry in investigations of integrability in AdS/CFT. (Readers are urged to refer to \cite{reviews} for reviews.)
It leads to bulk $S$-matrix \cite{Be} that obeys (twisted) Yang-Baxter equation (YBE). (Also refer to \cite{AFZ} for the corresponding bulk $S$-matrix that obeys standard YBE.) 
Such a $S$-matrix can be used to prove \cite{Be, MM, dL} a conjectured set of asymptotic Bethe equations \cite{BS} 
for the spectrum of gauge/string theory.  In connection to the open string/spin chain sector of 
AdS/CFT \cite{BV}-\cite{CY2}, these results have also been generalized to cases with boundaries, where the corresponding 
boundary $S$-matrices have been derived. Hofman and Maldacena \cite{HM} proposed boundary $S$-matrices corresponding to 
open strings attached to maximal giant gravitons \cite{giants} in $AdS_{5} \times S^{5}$ that describe the reflection of world-sheet
excitations for two cases: $Y = 0$ and $Z = 0$ giant graviton branes, hence generalizing the scattering theory of magnons in the planar  
limit of the AdS/CFT correspondence by including boundaries. In \cite{AN}, related boundary $S$-matrices (that indeed obey standard boundary Yang Baxter equation (BYBE) \cite{Ch, GZ}) 
were derived. Recently, transfer matrices for open chain for AdS/CFT were derived \cite{MN} and subsequently the corresponding all loop Bethe ansatz equations 
have been presented by Galleas in \cite{Galleas}.

A $q$-deformation of the above mentioned centrally-extended $su(2|2)$ algebra of AdS/CFT have been proposed by Beisert and Koroteev \cite{BK}.  
They derived the corresponding $q$-deformed bulk $S$-matrix, which they related to a deformation \cite{AB} of the 
one-dimensional Hubbard model \cite{EFGKK}. The related factorizable boundary $S$-matrices that obey the standard 
BYBE have been derived recently by using the Zamolodchikov-Fadeev (ZF)
algebra \cite{MuN}.  While derivation of all loop Bethe ansatz equations for this case is an interesting problem 
(analogous to that given by Galleas for the $su(2|2)$ case in \cite{Galleas}), construction of commuting open-chain/boundary transfer 
matrices (as observed by Sklyanin \cite{Sk}) is crucial for such a derivation. The fact that the $q$-deformed case of $SU(2|2)$ matrices is not as widely explored
in literature relative to the $q = 1$ case, has motivated us to consider this problem and present the material in a relatively more unified way.
The $q$-deformed bulk $S$-matrix is not of the difference form and has a peculiar crossing property \cite{BK}, thus motivating one to consider the generalization of 
Sklyanin's construction \cite{Sk}. Although such a construction for the undeformed $su(2|2)$ algebra was given in \cite{MN}, 
in this note, we consider an analogous construction for the $q$-deformed case, given that there might be details that deserve generalizations which could not have emerged
by considering the undeformed case alone. Indeed, we rely on the crossing property obeyed by the $q$-deformed bulk $S$-matrix to
show that the transfer matrix contains a crucial additional factor which is essential for commutativity. Such a factor
emerges for both the graded and non-graded versions of the $q$-deformed bulk $S$-matrices. These factors reduce to that obtained for the 
undeformed case as $q\rightarrow 1$ \cite{MN}.
 
The outline of this note is as follows. In Section \ref{sec:bulk}, we review the $q$-deformed bulk $S$-matrix that obeys standard YBE. 
Next, closely following the method outlined in \cite{AN}, we reformulate another derivation of the crossing symmetry for the $q$-deformed bulk $S$-matrix \cite{BK} in Section \ref{sec:crossing}. 
In Section \ref{sec:transfer}, we construct two different commuting open-chain transfer 
matrices, first with non-graded $q$-deformed bulk $S$-matrix and the second, with 
graded version of the corresponding bulk $S$-matrix. Next, we argue an extra factor is necessary for both versions in order for the transfer matrices to commute.
We conclude in Section \ref{sec:discuss} with a brief discussion of our results. 

\section{The $q$-deformed $SU(2|2)$-invariant bulk $S$-matrix}\label{sec:bulk}

We first briefly review the action of the symmetry generators (three Cartan generators $h_{j}$, three simple positive roots $E_{j}$ and three simple negative roots $F_{j}$, $j=1\,, 2\,, 3.$)
on the ZF operators, which we will denote by $A_{i}^{\dagger}(p)$, $i=1\,, 2\,, 3\,, 4$ following \cite{AFZ, AN}. The generators $E_{2}\,, F_{2}$ are fermionic, while the remaining ones are bosonic. (Readers are 
urged to refer to \cite{BK, MuN} for more detailed discussions.)   

\subsection{Bulk ZF algebra}\label{subsec:bulkZF} 

The action of the symmetry generators on the ZF operators can be obtained from the following two requirements: 
The one-particle states $A_{i}^{\dagger}(p) |0\rangle$, $|0\rangle$ being the vacuum state, must form a
fundamental representation of the symmetry algebra (see Eq.  (2.55) in
\cite{BK}); and multi-particle states must form higher (reducible) representations.  Finally, together with the fact
that the symmetry generators annihilate the vacuum state, the action of these generators on ZF operators is obtained. We review the results 
below which are reproduced from \cite{MuN} :
\newline \noindent A. The nontrivial commutators of the Cartan generators 
with the ZF operators are given by
\be
h_{1}\, A_{1}^{\dagger}(p) &=& -A_{1}^{\dagger}(p) +  
A_{1}^{\dagger}(p)\, h_{1}\,, \qquad
h_{1}\, A_{2}^{\dagger}(p) = A_{2}^{\dagger}(p) +  
A_{2}^{\dagger}(p)\, h_{1}\,,  \non \\
h_{3}\, A_{3}^{\dagger}(p) &=& -A_{3}^{\dagger}(p) +  
A_{3}^{\dagger}(p)\, h_{3}\,,  \qquad
h_{3}\, A_{4}^{\dagger}(p) = A_{4}^{\dagger}(p) +  
A_{4}^{\dagger}(p)\, h_{3}\,,   \label{CartanZF} \\
h_{2}\, A_{1}^{\dagger}(p) &=& -\big(C-\frac{1}{2}\big) A_{1}^{\dagger}(p) +  
A_{1}^{\dagger}(p)\, h_{2}\,, \qquad
h_{2}\, A_{2}^{\dagger}(p) = -\big(C+\frac{1}{2}\big) A_{2}^{\dagger}(p) +  
A_{2}^{\dagger}(p)\, h_{2}\,,  \non \\
h_{2}\, A_{3}^{\dagger}(p) &=& -\big(C-\frac{1}{2}\big) A_{3}^{\dagger}(p) +  
A_{3}^{\dagger}(p)\, h_{2}\,, \qquad
h_{2}\, A_{4}^{\dagger}(p) = -\big(C+\frac{1}{2}\big) A_{4}^{\dagger}(p) +  
A_{4}^{\dagger}(p)\, h_{2}\,,  \non 
\ee
$C=C(p)$ denotes the value of the corresponding central charge
\be
C &=& -\frac{1}{2}h_{1} -h_{2} -\frac{1}{2}h_{3}
\ee
The remaining such commutators 
are trivial, $h_{j}\, A_{k}^{\dagger}(p) =   A_{k}^{\dagger}(p)\, 
h_{j}$.

\noindent B. The nontrivial commutators of the bosonic simple roots with the ZF
operators are given by
\be
E_{1}\, A_{1}^{\dagger}(p) &=& q^{1/2}\, A_{2}^{\dagger}(p)\, 
q^{-h_{1}/2} +  q^{-1/2}\, A_{1}^{\dagger}(p)\, E_{1}\,, \qquad
E_{1}\, A_{2}^{\dagger}(p) = q^{1/2}\, A_{2}^{\dagger}(p)\, E_{1}\,, 
\non \\
E_{3}\, A_{4}^{\dagger}(p) &=& q^{-1/2}\, A_{3}^{\dagger}(p)\, 
q^{-h_{3}/2} +  q^{1/2}\, A_{4}^{\dagger}(p)\, E_{3}\,, \qquad
E_{3}\, A_{3}^{\dagger}(p) = q^{-1/2}\, A_{3}^{\dagger}(p)\, E_{3}\,, 
\non \\
F_{1}\, A_{2}^{\dagger}(p) &=& q^{-1/2}\, A_{1}^{\dagger}(p)\, 
q^{-h_{1}/2} +  q^{1/2}\, A_{2}^{\dagger}(p)\, F_{1}\,, \qquad
F_{1}\, A_{1}^{\dagger}(p) = q^{-1/2}\, A_{1}^{\dagger}(p)\, F_{1}\,, 
\non \\
F_{3}\, A_{3}^{\dagger}(p) &=& q^{1/2}\, A_{4}^{\dagger}(p)\, 
q^{-h_{3}/2} +  q^{-1/2}\, A_{3}^{\dagger}(p)\, F_{3}\,, \qquad
F_{3}\, A_{4}^{\dagger}(p) = q^{1/2}\, A_{4}^{\dagger}(p)\, F_{3}\,.
\label{bosonicsimpleZF}
\ee
The remaining such commutators are trivial,
\be
E_{1}\, A_{\alpha}^{\dagger}(p) &=& A_{\alpha}^{\dagger}(p)\, E_{1}\,, \qquad
F_{1}\, A_{\alpha}^{\dagger}(p) = A_{\alpha}^{\dagger}(p)\, F_{1}\,, \qquad
\alpha = 3, 4 \,, \non \\
E_{3}\, A_{a}^{\dagger}(p) &=& A_{a}^{\dagger}(p)\, E_{3}\,, \qquad
F_{3}\, A_{a}^{\dagger}(p) = A_{a}^{\dagger}(p)\, F_{3}\,, \qquad
a = 1, 2 \,.
\label{bs2}
\ee

\noindent C. Finally, the commutators of the fermionic generators with the ZF
operators are given by
\be
E_{2}\, A_{2}^{\dagger}(p) &=& e^{-i p/2}\left[
a(p)\, A_{4}^{\dagger}(p)\, 
q^{-h_{2}/2} +  q^{-(C+\frac{1}{2})/2}\, A_{2}^{\dagger}(p)\, E_{2} 
\right]\,, \non \\
E_{2}\, A_{3}^{\dagger}(p) &=& e^{-i p/2}\left[
b(p)\, A_{1}^{\dagger}(p)\, 
q^{-h_{2}/2} -  q^{-(C-\frac{1}{2})/2}\, A_{3}^{\dagger}(p)\, E_{2} 
\right]\,, \non \\
F_{2}\, A_{1}^{\dagger}(p) &=& e^{i p/2}\left[
c(p)\, A_{3}^{\dagger}(p)\, 
q^{-h_{2}/2} +  q^{-(C-\frac{1}{2})/2}\, A_{1}^{\dagger}(p)\, F_{2} 
\right]\,, \non \\
F_{2}\, A_{4}^{\dagger}(p) &=& e^{i p/2}\left[
d(p)\, A_{2}^{\dagger}(p)\, 
q^{-h_{2}/2} -  q^{-(C+\frac{1}{2})/2}\, A_{4}^{\dagger}(p)\, F_{2} 
\right]\,, 
\label{fermionicZF1}
\ee
and
\be
E_{2}\, A_{1}^{\dagger}(p) &=& e^{-i p/2} q^{-(C-\frac{1}{2})/2}\, 
A_{1}^{\dagger}(p)\, E_{2} \,,  \qquad 
E_{2}\, A_{4}^{\dagger}(p) = -e^{-i p/2} q^{-(C+\frac{1}{2})/2}\, 
A_{4}^{\dagger}(p)\, E_{2} \,, \non \\
F_{2}\, A_{2}^{\dagger}(p) &=& e^{i p/2} q^{-(C+\frac{1}{2})/2}\, 
A_{2}^{\dagger}(p)\, F_{2} \,,  \qquad 
F_{2}\, A_{3}^{\dagger}(p) = -e^{i p/2} q^{-(C-\frac{1}{2})/2}\, 
A_{3}^{\dagger}(p)\, F_{2} \,.
\label{fermionicZF2}
\ee

The functions $a(p)\,, b(p)\,, c(p)\,, d(p)$ are given below. (Refer  to \cite{BK} and \cite{MuN} for details used to set these functions.) 
\be
a &=& \sqrt{g}\, \gamma\, q^{-C} \,, \non \\
b &=& \sqrt{g}\, \frac{1}{x^{-}\gamma}
\left(x^{-} - q^{2C-1} x^{+}\right)  \,, \non \\
c &=& i\sqrt{g}\, \gamma\, \frac{q^{-C+\frac{1}{2}}}{x^{+}}\,, \non \\
d &=& i\sqrt{g}\,  \frac{q^{-\frac{1}{2}}}{\gamma}
\left( q^{2C+1} x^{-} -  x^{+}\right)\,, 
\label{abcd}
\ee 
with 
\be
e^{i p} = \frac{x^{+}}{q x^{-}} \,.
\label{momentum}
\ee
Together with (\ref{momentum}), one also need the following constraint to determine $x^{\pm}(p)$, 
\be
\frac{x^{+}}{q} +\frac{q}{x^{+}} - q x^{-} - \frac{1}{q x^{-}} 
+ i g (q - q^{-1})\left(\frac{x^{+}}{q x^{-}} -\frac{q 
x^{-}}{x^{+}}\right) = \frac{i}{g} \,.
\label{quadratic}
\ee 
As in \cite{BK}, we leave $\gamma$ unspecified at this point. We also recall that expressions for $a$ and $d$ in (\ref{abcd}) differ 
from those in \cite{BK} by factors of $q^{\mp C}$. 

\subsection{Nonzero matrix elements}

The two-particle $S$-matrix (up to a phase) is determined
by demanding that the symmetry generators commute with two-particle
scattering: from
$J\, A_{i}^{\dagger}(p_{1})\, A_{j}^{\dagger}(p_{2})
|0\rangle$ where $J$ is a symmetry generator, and assuming
that $J$ annihilates the vacuum state, one
arrives at linear combinations of $A_{j'}^{\dagger}(p_{2})\,
A_{i'}^{\dagger}(p_{1}) |0\rangle$ in two different ways, by applying the
ZF relation, 
\be
A_{i}^{\dagger}(p_{1})\, A_{j}^{\dagger}(p_{2}) = 
S_{i\, j}^{i' j'}(p_{1}, p_{2})\, 
A_{j'}^{\dagger}(p_{2})\, A_{i'}^{\dagger}(p_{1}) \,,
\label{bulkS1}
\ee
 and the symmetry relations 
(\ref{CartanZF}) - (\ref{fermionicZF2})
in different orders.  The consistency condition yields
a system of linear equations for the $S$-matrix
elements. The bulk $S$-matrix can then be defined by                                                                                                            
\be
S = S_{i\, j}^{i' j'} e_{i\, i'}\otimes e_{j\, j'}\,,
\label{bulkS2}
\ee
where $S_{i\, j}^{i' j'}$ represents the bulk $S$-matrix elements and summation over repeated indices is implied.
Noting that $e_{i j}$ is the usual elementary $4 \times 4$ matrix whose 
$(i, j)$ matrix element is 1, and all others are zero, (\ref{bulkS2}) represents the arrangement of the matrix elements into a $16 \times 16$ matrix.
Below, we list the nonzero matrix elements for the $q$-deformed bulk $S$-matrix, reproduced from \cite{MuN}, 
\be
S_{a\, a}^{a\, a} &=& \mathcal{ A}\,, \qquad \qquad \ \
S_{\alpha\, \alpha}^{\alpha\, \alpha} = \mathcal{ D}\,, \non \\
S_{a\, b}^{a\, b} &=& 
\frac{\mathcal{ A}-\mathcal{ B}}{q+q^{-1}}\,, \qquad 
S_{a\, b}^{b\, a} = \frac{q^{-\epsilon_{a b}}\mathcal{ A}
+q^{\epsilon_{a b}}\mathcal{ B}}{q+q^{-1}} \,, \non \\
S_{\alpha\, \beta}^{\alpha\, \beta} &=& 
\frac{\mathcal{ D}-\mathcal{ E}}{q+q^{-1}}\,, \quad 
S_{\alpha\, \beta}^{\beta\, \alpha} = 
\frac{q^{-\epsilon_{\alpha \beta}}\mathcal{ D}
+q^{\epsilon_{\alpha \beta}}\mathcal{ E}}{q+q^{-1}} \,, \non \\
S_{a\, b}^{\alpha\, \beta} &=& 
q^{(\epsilon_{a b}-\epsilon_{\alpha \beta})/2}
\epsilon_{a b}\epsilon^{\alpha \beta}\, \frac{\mathcal{ C}}{q+q^{-1}} \,, \quad
S_{\alpha\, \beta}^{a\, b} = 
q^{(\epsilon_{\alpha \beta}-\epsilon_{a b})/2}
\epsilon^{a b}\epsilon_{\alpha \beta}\, \frac{\mathcal{ F}}{q+q^{-1}} \,, \non \\
S_{a\, \alpha}^{a\, \alpha} &=& \mathcal{ L}\,, \quad 
S_{a\, \alpha}^{\alpha\, a} = \mathcal{ K} \,, \quad 
S_{\alpha\, a}^{a\, \alpha} = \mathcal{ H}\,, \quad 
S_{\alpha\, a}^{\alpha\, a} = \mathcal{ G} \,,  
\label{bulkS3}
\ee
where $a\,, b \in \{1\,, 2\}$ with $a \ne b$;  
$\alpha\,, \beta \in \{3\,, 4\}$ with $\alpha \ne \beta$; and
\be
\mathcal{ A} &=& A^{BK}_{21} = 
S_{0}q^{C_{2}-C_{1}}e^{i(p_{2}-p_{1})/2}
\frac{x^{+}_{1}-x^{-}_{2}}{x^{-}_{1}-x^{+}_{2}}\,, \non \\
\mathcal{ B} &=& B^{BK}_{21} = 
S_{0}q^{C_{2}-C_{1}}e^{i(p_{2}-p_{1})/2}\frac{x^{+}_{1}-x^{-}_{2}}{x^{-}_{1}-x^{+}_{2}}
\left(1-(q+q^{-1})q^{-1}\frac{x^{+}_{1}-x^{+}_{2}}{x^{+}_{1}-x^{-}_{2}}
\frac{x^{-}_{1}-s(x^{+}_{2})}{x^{-}_{1}-s(x^{-}_{2})}\right)
\,, \non \\
\mathcal{ C} &=& q^{-(C_{1}+C_{2}-1)/2}C^{BK}_{21} = S_{0}(q+q^{-1})
i g q^{(C_{2}-5C_{1}-2)/2}e^{i (p_{2}-2p_{1})/2} \gamma_{1}\gamma_{2}
\frac{i g^{-1}x^{+}_{1}-(q-q^{-1})}{x^{-}_{1}-s(x^{-}_{2})}\non \\
& & \cdot \frac{s(x^{+}_{1})-s(x^{+}_{2})}{x^{-}_{1}-x^{+}_{2}}
\,, \non
\ee
\be 
\mathcal{ D} &=& -S_{0}\,, \non \\
\mathcal{ E} &=& E^{BK}_{21} = -S_{0}\left(1-(q+q^{-1})q^{-2C_{1}-1}e^{-i p_{1}}
\frac{x^{+}_{1}-x^{+}_{2}}{x^{-}_{1}-x^{+}_{2}}
\frac{x^{+}_{1}-s(x^{-}_{2})}{x^{-}_{1}-s(x^{-}_{2})}\right)
\,, \non \\
\mathcal{ F} &=& q^{(C_{1}+C_{2}-1)/2}F^{BK}_{21} = -S_{0}(q+q^{-1})
i g q^{(5C_{2}-C_{1}-2)/2}e^{i (2p_{2}-p_{1})/2}
\frac{i g^{-1}x^{+}_{1}-(q-q^{-1})}{x^{-}_{1}-s(x^{-}_{2})}
\frac{s(x^{+}_{1})-s(x^{+}_{2})}{x^{-}_{1}-x^{+}_{2}}\non \\
& & \cdot \frac{1}{1-g^{2}(q-q^{-1})^{2}} \frac{1}{\gamma_{1}\gamma_{2}}
(x^{+}_{1}-x^{-}_{1})(x^{+}_{2}-x^{-}_{2})
\,, \non \\
\mathcal{ G} &=& G^{BK}_{21} = S_{0}q^{-C_{1}-1/2} e^{-i p_{1}/2} 
\frac{x^{+}_{1}-x^{+}_{2}}{x^{-}_{1}-x^{+}_{2}}
\,, \non \\
\mathcal{ H} &=& q^{(C_{1}-C_{2})/2}H^{BK}_{21} = S_{0}q^{(C_{1}-C_{2})/2}
\frac{\gamma_{2}}{\gamma_{1}}
\frac{x^{+}_{1}-x^{-}_{1}}{x^{-}_{1}-x^{+}_{2}}
\,, \non \\
\mathcal{ K} &=& q^{-(C_{1}-C_{2})/2}K^{BK}_{21} = S_{0}q^{3(C_{2}-C_{1})/2}
e^{i(p_{2}-p_{1})/2}\frac{\gamma_{1}}{\gamma_{2}}
\frac{x^{+}_{2}-x^{-}_{2}}{x^{-}_{1}-x^{+}_{2}}
\,, \non \\
\mathcal{ L} &=& L^{BK}_{21}= S_{0}q^{C_{2}+1/2} e^{i p_{2}/2}
\frac{x^{-}_{1}-x^{-}_{2}}{x^{-}_{1}-x^{+}_{2}}
\,, \label{bulkS4}
\ee 
where $A^{BK}_{21}\,, B^{BK}_{21}\,, \ldots$ denote the amplitudes
$A_{12}\,, B_{12}\,, \ldots$ in Table 2 of \cite{BK}, respectively,
with labels 1 and 2 interchanged.  $S_{0}$ is the overall scalar
factor (denoted by $R^{0}$ in \cite{BK}).  The function
$s(x)$ is the ``antipode map'' defined by \cite{BK}
\be
s(x) = \frac{1 - ig (q-q^{-1}) x}{x + ig (q-q^{-1})} \,,
\label{sfunc}
\ee
which has the limit $s(x) \rightarrow 1/x$ for $q \rightarrow 1$. Furthermore, $C_{1}\equiv C(p_{1})\,, C_{2}\equiv C(p_{2})$ are determined from \cite{BK}
\be
q^{2C} = \frac{1}{q} \left(\frac{1 - i g (q - q^{-1}) x^{+}}
{1 - i g (q - q^{-1}) x^{-}} \right) = 
q  \left(\frac{1 +  i g (q - q^{-1})/x^{+}}
{1 + i g (q - q^{-1})/x^{-}} \right) \,.
\label{qC}
\ee
As pointed out in \cite{MuN}, the
amplitudes $\mathcal{ C}$, $\mathcal{ F}$, $\mathcal{ H}$ and
$\mathcal{ K}$ have extra factors involving powers of $q$
with respect to the amplitudes given in \cite{BK}.  Nevertheless, it has been verified that the above given $S$-matrix still satisfies the
(standard) Yang-Baxter equation
\be
S_{12}(p_{1}, p_{2})\, S_{13}(p_{1}, p_{3})\, S_{23}(p_{2}, p_{3})\ =
S_{23}(p_{2}, p_{3})\, S_{13}(p_{1}, p_{3})\, S_{12}(p_{1}, p_{2}) \,.
\label{YBE}
\ee
 even without those extra factors. We use the standard convention $S_{12} = S \otimes
\id$, $S_{23} = \id \otimes S$, and $S_{13} = {\cal P}_{12}\, S_{23}\,
{\cal P}_{12}$, where ${\cal P}_{12} = {\cal P} \otimes \id$, ${\cal
P}$ is the permutation matrix, and $\id$ is the four-dimensional
identity matrix.  In addition, as for the undeformed ($q=1$) matrix, the $q$-deformed matrix has the following unitarity property
\be
S_{12}(p_{1}, p_{2})\, S_{21}(p_{2}, p_{1}) = \id \,,
\label{unitarity}
\ee
provided that the bulk scalar factor obeys
\be
S_{0}(p_{1},p_{2})S_{0}(p_{2}, p_{1}) = 1
\ee
where $S_{21} = {\cal P}_{12}\, S_{12}\, {\cal P}_{12}$, as well as 
the crossing property \cite{BK} which we shall describe in more detail in the following Section.

\section{Crossing symmetry}\label{sec:crossing}

In this Section, we reformulate the derivation of crossing equation for the $q$-deformed bulk $S$-matrix 
given in \cite{BK}, following closely the method outlined in \cite{AN} in terms of ZF operators.
This property is needed to construct the commuting open-chain transfer matrices.
Following \cite{AN}, we begin by defining the ``singlet'' operator
\be
I(p) = \mathrm{C}^{i j}(p)\, A^{\dagger}_{i}(p)\, A^{\dagger}_{j}(\bar 
p) & \equiv & 
\alpha(p)\,  A^{\dagger}_{1}(p)\, A^{\dagger}_{2}(\bar 
p) + \beta(p)\,  A^{\dagger}_{2}(p)\, A^{\dagger}_{1}(\bar 
p) + \sqrt{q} A^{\dagger}_{3}(p)\, 
A^{\dagger}_{4}(\bar p) \non \\
&-& {1\over \sqrt{q}} A^{\dagger}_{4}(p)\, 
A^{\dagger}_{3}(\bar p) \,,
\label{singlet}
\ee
where the functions $\alpha(p)\,, \beta(p)$ are yet to be determined. 
Hence, $\mathrm{C}(p)$ is the $4 \times 4$ matrix
\be
\mathrm{C}(p)= \left( \begin{array}{cccc}
0  &\alpha(p)  & 0 & 0 \\
\beta(p) &0  &0 & 0 \\
0  &0  &0 & \sqrt{q} \\
0  &0  &-{1\over \sqrt{q}} & 0
\end{array} \right) \,.
\label{CCmatrix}
\ee 
$\bar p = -p$ denotes the antiparticle momentum, with \cite{BK} 
\be
x^{\pm}(\bar p) = s(x^{\pm}(p)) \,,
\label{barp}
\ee
Functions $\alpha(p)\,, \beta(p)$ are determined by the conditions that the singlet 
operator commutes with the fermionic generators. Indeed, using (\ref{fermionicZF1}) and (\ref{fermionicZF2}),
the condition $E_{2}I(p) |0\rangle = I(p)\, E_{2} |0\rangle =0$ leads to  \footnote{Similar matrix is given in \cite{BK}}
\be
\alpha(p) = - e^{-i p/2} \sqrt{q} \frac{b(p)}{a(\bar p)}= e^{i p/2} \sqrt{q} \frac{b(\bar p)}{a(p)}= i \sqrt{q}\sgn (p)\,, \non\\ 
\beta(p) = - e^{i p/2} {1\over \sqrt{q}} \frac{b(\bar p)}{a(p)}= e^{-i p/2} {1\over \sqrt{q}} \frac{b(p)}{a(\bar p)}= -i{1\over \sqrt{q}}\sgn (p) \,.
\label{Celements}
\ee
We remark that (\ref{CCmatrix}) reduces to the matrix $\mathrm{C}(p)$ in \cite{AN} as $q\rightarrow 1$. We also note the following property of the matrix $\mathrm{C}(p)$,
\be
\mathrm{C}(-p) = - \mathrm{C}(p)^{-1}
\label{Cprop}
\ee
which is needed to construct the desired transfer matrix.
A particular choice for $\gamma(p)$ that evidently appears in $a(p)\,, b(p)\,, c(p)\,, d(p)$ as given in (\ref{abcd}) is
\be
\gamma(p) = {\sqrt{-q^{C}e^{i p/2}(x^{+}(p) - x^{-}(p))}\over \sqrt [4]{1 - (q - q^{-1})^{2} g^{2}}}
\label{gamm1}
\ee
which can be used to verify the crossing property that follows, namely (\ref{cross1}) and (\ref{cross2}) below. $C$ in (\ref{gamm1}) again refers to the value of central charge. 
The expression for $\gamma$ used here is essentially the same as (2.65) in \cite{BK} up to a certain constant. This difference is presumably due to the fact 
that our expressions for $a$ and $d$ differ  from those in \cite{BK} by factors of $q^{\mp C}$.\footnote
{One could also use the relation \cite{BK} $\gamma(p)\gamma(\bar p) = -i(q^{C}e^{i p/2} -  q^{-C}e^{-i p/2})$ to verify crossing property (\ref{cross1}) and (\ref{cross2}). 
Again, this expression is essentially the same as in \cite{BK} up to a certain constant presumably due to the difference in our expressions for $a$ and $d$ compared to those in \cite{BK}.} 
With regard to crossing property (\ref{cross1}) and (\ref{cross2}), (\ref{gamm1}) is also consistent with the bulk $S$-matrix presented in the last Section where the amplitudes $\mathcal{C}\,, \mathcal{F}\,, \mathcal{H}\,, \mathcal{K}$ have 
extra factors involving powers of $q$.  As noticed in \cite{BK}, (\ref{gamm1}) possesses nice properties analogous to that of the undeformed case \cite{AFr}. 
Having found the matrix $\mathrm{C}(p)$ (as given by (\ref{CCmatrix}) and (\ref{Celements})), we now begin the reformulation of crossing property for the $q$-deformed bulk $S$-matrix
following closely the method outlined in \cite{AN} for the $q = 1$ case. As pointed out in \cite{AN} for $q=1$ case, the requirement that the singlet 
operator scatter trivially with a particle, along with (\ref{bulkS1}) and (\ref{singlet}) lead to the following,
\be
A^{\dagger}_{i}(p_{1})\, I(p_{2}) &=&
\mathrm{C}^{j k}(p_{2})\, A^{\dagger}_{i}(p_{1})\, A^{\dagger}_{j}(p_{2})\, 
A^{\dagger}_{k}(\bar p_{2})\non\\
&=& \mathrm{C}^{j k}(p_{2})\, S_{i j}^{i' j'}(p_{1},p_{2})\, A^{\dagger}_{j'}(p_{2})\,
A^{\dagger}_{i'}(p_{1})\, A^{\dagger}_{k}(\bar p_{2})\non\\
&=& \mathrm{C}^{j k}(p_{2})\, S_{i j}^{i' j'}(p_{1},p_{2})\, S_{i' k}^{i'' k'}(p_{1}, \bar p_{2})\, 
A^{\dagger}_{j'}(p_{2})\, A^{\dagger}_{k'}(\bar p_{2})\, A^{\dagger}_{i''}(p_{1}) \non\\
&\equiv& I(p_{2})\, A^{\dagger}_{i}(p_{1})
\ee
that implies
\be
\mathrm{C}^{j k}(p_{2})\, S_{i j}^{i' j'}(p_{1},p_{2})\, S_{i' k}^{i'' k'}(p_{1}, \bar p_{2}) =
\mathrm{C}^{j' k'}(p_{2})\, \delta_{i}^{i''} \,,
\ee
One can re-write the above equation in matrix notation as
\be
S_{12}^{t_{2}}(p_{1},p_{2})\, \mathrm{C}_{2}(p_{2})\, S_{12}(p_{1}, \bar p_{2})\, 
\mathrm{C}_{2}(p_{2})^{-1}  =  \id\,.
\label{cross1}
\ee 
which is the desired crossing property for the $q$-deformed bulk $S$-matrix. 
The following equivalent form of (\ref{cross1}) can be obtained by applying the permutation and exchanging $p_{1}$ and $p_{2}$,
\be
S_{21}^{t_{1}}(p_{2},p_{1})\, \mathrm{C}_{1}(p_{1})\, S_{21}(p_{2}, \bar p_{1})\, 
\mathrm{C}_{1}(p_{1})^{-1}  =  \id\,.
\label{cross2}
\ee 
In (\ref{cross1}) and (\ref{cross2}) above, $\mathrm{C}_{1} = \mathrm{C} \otimes
\id$, $\mathrm{C}_{2} = \id \otimes \mathrm{C}$; $t_{1}$ and $t_{2}$ are the transposition in the first and the second space respectively.  
Using (\ref{bulkS3}), (\ref{bulkS4}), (\ref{CCmatrix}) and (\ref{Celements}) in (\ref{cross1}), one obtains,
\be
S_{0}(p_{1},p_{2})S_{0}(p_{1},\bar p_{2}) = {1\over f(p_{1},p_{2})}
\label{crossrel}
\ee
for the bulk scalar factor where \cite{BK}\footnote{Refer to \cite{BK} for a number of other equivalent forms for $f(p_{1},p_{2})$}
\be
f(p_{1},p_{2}) ={1\over q} \frac{\left(s(x^{+}_{1}) - 
x^{-}_{2}\right)(x^{+}_{1} - x^{+}_{2})}
{\left(s(x^{-}_{1}) - 
x^{-}_{2}\right)(x^{-}_{1} - x^{+}_{2})} \,.
\label{ffunc}
\ee
It would be interesting to find a solution of (\ref{crossrel}).

\section{Transfer matrix}\label{sec:transfer}

In this section, we present the Sklyanin's construction (\cite{Sk}) of the open chain transfer matrix.
Bulk and boundary $S$-matrices are the two main building blocks of the
transfer matrix.  While the $q$-deformed bulk $S$-matrix is as given in (\ref{bulkS3}) and (\ref{bulkS4}) which obeys
the standard YBE (\ref{YBE}),  the $q$-deformed right boundary $S$-matrix $R^{-}(p)$ is a diagonal matrix
found in \cite{MuN} for the $Y=0$ giant graviton brane, 
\be
R^{-}(p) =  \diag( 
-\frac{e^{i p}}{x^{+} s(x^{-})}\frac{\gamma(p)}{\gamma(-p)} \,,
e^{-ip}\frac{\gamma(p)}{\gamma(-p)}\,,  
1 \,, 1 ) \,.
\label{boundarySright}
\ee
in a basis where the standard (right) boundary Yang-Baxter equation (BYBE) \cite{Ch,
GZ} 
\be
S_{12}(p_{1}, p_{2})\, R_{1}^{-}(p_{1})\, S_{21}(p_{2}, -p_{1})\, 
R_{2}^{-}(p_{2}) = 
R_{2}^{-}(p_{2})\, S_{12}(p_{1}, -p_{2})\, R_{1}^{-}(p_{1})\, 
S_{21}(-p_{2}, -p_{1}) 
\label{BYBE}
\ee 
is satisfied. Note that in the $q\rightarrow 1$ limit, (\ref{boundarySright}) reduces to the corresponding 
undeformed boundary $S$-matrix in (\cite{AN}). We also recall from \cite{MuN}, 
\be
x^{\pm}(-p) =  -{1\over s(x^{\mp}(p))} 
\label{negation}
\ee 
which is crucial to study boundary scattering.
The following monodromy matrices can then be constructed from the bulk $S$-matrix,
\be
T_{a}(p\,; \{p_{i}\}) &=& S_{a N}(p,p_{N}) \cdots S_{a 1}(p,p_{1}) \,,
\non \\
\widehat T_{a}(p\,; \{p_{i}\}) &=& S_{1 a}(p_{1}, -p) \cdots S_{N 
a}(p_{N}, -p) \,,
\label{monodromy}
\ee
where $\{p_{1}, \ldots, p_{N}\}$ are arbitrary ``inhomogeneities''
associated with each of the $N$ quantum spaces, and the auxiliary
space is denoted by $a$. The quantum-space ``indices'' are
suppressed from the monodromy matrices. Further,  the ``decorated'' right boundary $S$-matrix given by
\be
{\cal T}^{-}_{a}(p\,; \{p_{i}\}) = T_{a}(p\,; \{p_{i}\})\,
R_{a}^{-}(p)\,  \widehat T_{a}(p\,; \{p_{i}\})
\ee
also satisfies the BYBE, i.e.,
\be
\lefteqn{S_{a b}(p_{a}, p_{b})\, {\cal T}_{a}^{-}(p_{a}\,; \{p_{i}\})\, 
S_{b a}(p_{b}, -p_{a})\, 
{\cal T}_{b}^{-}(p_{b}\,; \{p_{i}\})} \non \\
& &= {\cal T}_{b}^{-}(p_{b}\,; \{p_{i}\})\, S_{a b}(p_{a}, -p_{b})\, {\cal 
T}_{a}^{-}(p_{a}\,; \{p_{i}\})\, 
S_{b a}(-p_{b}, -p_{a}) \,,
\ee 
 because of (\ref{BYBE}) and the following relations obeyed by the monodromy matrices,
\be
S_{a b}(p_{a}, p_{b})\, T_{a}(p_{a}\,; \{p_{i}\})\,
T_{b}(p_{b}\,; \{p_{i}\})  &=& T_{b}(p_{b}\,; \{p_{i}\})\,
T_{a}(p_{a}\,; \{p_{i}\})\, S_{a b}(p_{a}, p_{b}) \,, \non \\
S_{b a}(-p_{b}, -p_{a})\, \widehat T_{a}(p_{a}\,; \{p_{i}\})\,
\widehat T_{b}(p_{b}\,; \{p_{i}\}) &=& 
\widehat T_{b}(p_{b}\,; \{p_{i}\})\, 
\widehat T_{a}(p_{a}\,; \{p_{i}\})\, S_{b a}(-p_{b}, -p_{a})\,, \non \\
\widehat T_{a}(p_{a}\,; \{p_{i}\})\, S_{b a}(p_{b}, -p_{a})  
T_{b}(p_{b}\,; \{p_{i}\}) &=& 
T_{b}(p_{b}\,; \{p_{i}\})\, S_{b a}(p_{b}, -p_{a})\, 
\widehat T_{a}(p_{a}\,; \{p_{i}\}) 
\label{fundamental}
\ee

Following Sklyanin \cite{Sk}, we assume that the open-chain transfer 
matrix is of the double-row form
\be
t(p\,; \{p_{i}\}) &=& \tr_{a} R_{a}^{+}(p)\, {\cal T}^{-}_{a}(p\,; 
\{p_{i}\})  \non \\
&=& \tr_{a} R_{a}^{+}(p)\, T_{a}(p\,; \{p_{i}\})\,
R_{a}^{-}(p)\,  \widehat T_{a}(p\,; \{p_{i}\}) \,,
\label{transfer}
\ee 
where the trace is taken over the auxiliary space, and the left boundary
$S$-matrix $R^{+}(p)$ is chosen to ensure the essential commutativity
property
\be
\left[ t(p\,; \{p_{i}\}) \,, t(p'\,; \{p_{i}\}) \right] = 0 
\label{commutativity}
\ee
for arbitrary values of $p$ and $p'$.  Making use of the unitarity and
crossing properties (\ref{unitarity}), (\ref{cross1}) and (\ref{cross2}), we find
that the commutativity property is indeed obeyed, provided that $R^{+}(p)$ obeys \footnote{Equivalent relation appears in \cite{MN}.}
\be
\lefteqn{S_{2 1}(p_{2}, p_{1})^{{t}_{12}}\, R^{+}_{1}(p_{1})^{{t}_{1}}\, 
\mathrm{C}_{1}(p_{1})^{-1}\, S_{21}(p_{2}, \overline{-p_{1}})^{t_{2}}\, 
\mathrm{C}_{1}(p_{1})\,  R^{+}_{2}(p_{2})^{{t}_{2}}} \non \\
& & =  R^{+}_{2}(p_{2})^{{t}_{2}}\, \mathrm{C}_{2}(p_{2})^{-1}\, 
S_{12}(p_{1}, \overline{-p_{2}})^{t_{1}}\, \mathrm{C}_{2}(p_{2})\,
R^{+}_{1}(p_{1})^{{t}_{1}}\, S_{1 2}(-p_{1}, -p_{2})^{{t}_{12}} \,.
\label{ugly}
\ee
where as defined in (\ref{barp}), $\bar p$ denotes the antiparticle momentum.
In obtaining this result, we also make use of (\ref{Cprop}) and the following identity
\be
f(p_{1},p_{2}) = f(-p_{2}, -p_{1}) 
\ee
for the function defined in (\ref{ffunc}).  Using (\ref{cross1}) and (\ref{cross2}), (\ref{ugly}) 
can be simplified to yield
\be
\lefteqn{S_{12}(p_{1}, p_{2})\, M_{1}^{-1}\, R_{1}^{+}(-p_{1})\, S_{21}(p_{2}, -p_{1})\, 
M_{2}^{-1}\, R_{2}^{+}(-p_{2})} \non \\
& & = M_{2}^{-1}\, R_{2}^{+}(-p_{2})\,  S_{12}(p_{1}, -p_{2})\, M_{1}^{-1}\, R_{1}^{+}(-p_{1})\,  
S_{21}(-p_{2}, -p_{1}) \,,
\label{good}
\ee 
where the matrix $M$ is given by
\be
M = \mathrm{C}(p)^{t}\, \mathrm{C}(p) = \diag \left(-1/q\,, -q\,, 1/q\,, q \right) \,.
\label{emm}
\ee
where $\mathrm{C}(p)^{t}$ is the transpose of $\mathrm{C}(p)$. In obtaining (\ref{good}), we make use of the identities
\be
f(p_{1},p_{2}) = f(\overline{-p_{2}}, \overline{-p_{1}}) 
\ee
and
\be
M_{1}\, S_{12}(p_{1}, p_{2})\, M_{2}^{-1} = M_{2}^{-1}\, S_{12}(p_{1}, p_{2})\, M_{1}\,.
\ee 
or equivalently
\be
M_{1}^{-1}\, S_{12}(p_{1}, p_{2})\, M_{2} = M_{2}\, S_{12}(p_{1}, p_{2})\, M_{1}^{-1}\,.
\ee 

Comparing the $R^{+}(p)$ relation (\ref{good}) with the $R^{-}(p)$ 
relation (\ref{BYBE}), we conclude that the left boundary $S$-matrix 
is given by
\be
R^{+}(p) =  M R^{-}(-p) \,,
\label{boundarySleft}
\ee
where $M$ is given by (\ref{emm}). We emphasize that this matrix $M$
is essential in order for the transfer matrix (\ref{transfer}) to have the commutativity property 
(\ref{commutativity}).\footnote{Similar matrices also appear in the construction 
of open-chain transfer matrices in \cite{BGZZ} and \cite{MezN}.}
We have verified (\ref{commutativity}) numerically for small 
numbers of sites. 

We also find that if we work instead with corresponding 
graded quantities \footnote{See for example \cite{graded} and \cite{BGZZ}.}
with the following parity assignments  
\be
p(1) = p(2) = 0\,, \qquad p(3) = p(4) = 1 \,,
\ee
and define the graded bulk $S$-matrix by (see, e.g., \cite{MM})
\be
S^{g}(p_{1}, p_{2}) = {\cal P}^{g}\, {\cal P}\, S(p_{1}, p_{2})\,,
\label{gradedS}
\ee
where ${\cal P}^{g}$ is the graded permutation matrix
\be
{\cal P}^{g} = \sum_{i,j=1}^{4} (-1)^{p(i) p(j)} e_{i\, j} \otimes 
e_{j\, i} \,,
\ee
the transfer matrix 
\be
t(p\,; \{p_{i}\}) = \str_{a}R_{a}^{+}(p)\, T_{a}(p\,; \{p_{i}\})\,
R_{a}^{-}(p)\,  \widehat T_{a}(p\,; \{p_{i}\}) \,.
\label{gradedtransfer2}
\ee
satisfies the commutativity property (\ref{commutativity}) provided $R^{+}(p)$ is given by
(\ref{boundarySleft}) with $M = \diag \left(1/q\,, q\,, 1/q\,, q \right) $, namely
\be
t(p\,; \{p_{i}\}) = \str_{a} M R_{a}^{-}(-p)\, T_{a}(p\,; \{p_{i}\})\,
R_{a}^{-}(p)\,  \widehat T_{a}(p\,; \{p_{i}\}) \,.
\label{gradedtransfer2}
\ee
In (\ref{gradedtransfer2}), $\str$ denotes the supertrace, the monodromy matrices are formed
as in (\ref{monodromy}) except with the graded $S$-matrix
(\ref{gradedS}) using the graded tensor product (instead of the
ordinary tensor product), and $R^{-}(p)$ is again given by
(\ref{boundarySright}), which also satisfies the graded BYBE.  
We have again numerically verified (\ref{commutativity}) for (\ref{gradedtransfer2}) with Mathematica for small 
numbers of sites.

\section{Discussion}\label{sec:discuss}

We have presented a commuting open-chain transfer
matrix given by (\ref{transfer}) constructed from the $q$-deformed $SU(2|2)$ bulk and 
boundary $S$-matrices, where $T_{a}(p\,; \{p_{i}\})$ and
$\widehat T_{a}(p\,; \{p_{i}\})$ are given by (\ref{monodromy}), and
$R^{+}(p)$ is given by (\ref{boundarySleft}), which contains the 
factor $M$ (\ref{emm}).  Alternatively, using graded 
version of the bulk $S$-matrix, we also constructed a transfer 
matrix (\ref{gradedtransfer2}) which still seems to include similar extra factor (unlike 
the undeformed case where such a factor can be avoided using graded versions of the $S$-matrices). 
These transfer matrices reduce to that obtained for the undeformed case when $q\rightarrow 1$ \cite{AN}. 

An interesting problem is to solve the open versions of the deformed Hubbard models based on 
the deformed boundary $S$-matrix and the corresponding transfer matrix. It will also be interesting 
to construct corresponding open chain transfer matrices that exclude 
the extra factor $M$. Perhaps such a construction will be more convenient for the formulation of the Bethe-Yang equation
on an interval with boundaries. We hope to be able to address these issues in the future.
 
\section*{Acknowledgments}
I would like to thank R. I. Nepomechie for his collaboration on earlier related projects.


\begin{thebibliography}{99}

\bibitem{Be}
N. Beisert,
``The $su(2|2)$ dynamic $S$-matrix,''
{\it Adv.Theor.Math.Phys.} {\bf 12}, 945 (2008) 

[arXiv:hep-th/0511082];\\
N. Beisert,
``The Analytic Bethe Ansatz for a Chain with Centrally Extended 
$su(2|2)$ Symmetry,''
{\it J. Stat. Mech.}  {\bf 0701}, P017 (2007)
[arXiv:nlin/0610017].

\bibitem{AFPZ}
G. Arutyunov, S. Frolov, J. Plefka and M. Zamaklar,
``The off-shell symmetry algebra of the light-cone $AdS_{5} \times S^{5}$
superstring,''
{\it J. Phys.} {\bf A40}, 3583 (2007)
[arXiv:hep-th/0609157].

\bibitem{reviews}
A.A. Tseytlin,
``Spinning strings and AdS/CFT duality,''
in Ian Kogan Memorial Volume, {\it From Fields to Strings: Circumnavigating
Theoretical Physics}, M. Shifman, A. Vainshtein, and J. Wheater, eds.
(World Scientific, 2004)
[arXiv:hep-th/0311139];\\ 
N. Beisert,
``The dilatation operator of ${\cal N} = 4$ super Yang-Mills theory and
integrability,''
{\it Phys. Rept.}  {\bf 405}, 1 (2005)
[arXiv:hep-th/0407277];\\
K. Zarembo,
``Semiclassical Bethe ansatz and AdS/CFT,''
{\it Comptes Rendus Physique} {\bf 5}, 1081 (2004)
[{\it Fortsch. Phys.}  {\bf 53}, 647 (2005)]
[arXiv:hep-th/0411191];\\
J. Plefka,
``Spinning strings and integrable spin chains in the AdS/CFT
correspondence,''
{\it Living Rev. Rel.}  {\bf 8}, 9 (2005)
[arXiv:hep-th/0507136];\\
J.A. Minahan, 
``A brief introduction to the Bethe ansatz in ${\cal N}=4$ 
super-Yang-Mills,''
{\it J. Phys.} {\bf A39}, 12657 (2006);\\
K. Okamura,
``Aspects of Integrability in AdS/CFT Duality,''
[arXiv:0803.3999].

\bibitem{AFZ}
G. Arutyunov, S. Frolov and M. Zamaklar,
`The Zamolodchikov-Faddeev algebra for $AdS_{5} \times S^{5}$ superstring,''
{\it JHEP} {\bf 0704}, 002 (2007)
[arXiv:hep-th/0612229].

\bibitem{MM}
M.J. Martins and C.S. Melo,
``The Bethe ansatz approach for factorizable centrally extended 
$S$-matrices,''
{\it Nucl. Phys.} {\bf B785}, 246 (2007) 
[arXiv:hep-th/0703086].

\bibitem{dL}
M. de Leeuw,
``Coordinate Bethe Ansatz for the String $S$-Matrix,''
{\it J. Phys.} {\bf A40}, 14413 (2007)
[arXiv:0705.2369].

\bibitem{BS}
N. Beisert and M. Staudacher,
``Long-range $PSU(2,2|4)$ Bethe ansaetze for gauge theory and strings,''
{\it Nucl. Phys.} {\bf B727}, 1 (2005)
[arXiv:hep-th/0504190].

\bibitem{BV}
D. Berenstein and S.E. Vazquez,
``Integrable open spin chains from giant gravitons,''
{\it JHEP} {\bf 0506}, 059 (2005)
[arXiv:hep-th/0501078].

\bibitem{MS}
T. McLoughlin and I. Swanson,
``Open string integrability and AdS/CFT,''
{\it  Nucl. Phys.} {\bf B723}, 132 (2005)
[arXiv:hep-th/0504203].

\bibitem{Ag}
A. Agarwal,
``Open spin chains in super Yang-Mills at higher loops: Some potential
problems with integrability,''
{\it JHEP} {\bf 0608}, 027 (2006)
[arXiv:hep-th/0603067];\\
K. Okamura and K. Yoshida,
``Higher loop Bethe ansatz for open spin-chains in AdS/CFT,''
{\it JHEP} {\bf 0609}, 081 (2006)
[arXiv:hep-th/0604100].

\bibitem{MV}
N. Mann and S.E. Vazquez,
``Classical open string integrability,''
{\it JHEP} {\bf 0704}, 065 (2007)
[arXiv:hep-th/0612038].

\bibitem{HM}
D.M. Hofman and J.M. Maldacena,
``Reflecting magnons,''
{\it JHEP} {\bf 0711}, 063 (2007)
[arXiv:0708.2272].

\bibitem{CC}
H.Y. Chen and D.H. Correa,
``Comments on the Boundary Scattering Phase,''
{\it JHEP} {\bf 0802}, 028 (2008)
[arXiv:0712.1361].

\bibitem{ABR}
C. Ahn, D. Bak and S.J. Rey,
``Reflecting Magnon Bound States,''
{\it JHEP} {\bf 0804}, 050 (2008)
[arXiv:0712.4144].

\bibitem{AN}
C. Ahn and R.I. Nepomechie,
``The Zamolodchikov-Faddeev algebra for open strings attached to giant
gravitons,''
{\it JHEP} {\bf 0805}, 059 (2008)
[arXiv:0804.4036].

\bibitem{BL}
N. Beisert and F. Loebbert,
``Open Perturbatively Long-Range Integrable $gl(N)$ Spin Chains,''
{\it Adv. Sci. Lett.} {\bf 2}, 261 (2009) 
[arXiv:0805.3260]. 

\bibitem{Pa}
L. Palla,
``Issues on magnon reflection,''
{\it Nucl.Phys.} {\bf B808}, 205 (2009) 
[arXiv:0807.3646].

\bibitem{CY}
D.H. Correa and C.A.S. Young,
``Reflecting magnons from D7 and D5 branes,''
{\it J.Phys.} {\bf A41}, 455401 (2008) 
[arXiv:0808.0452].

\bibitem{MN}
R. Murgan and R. I. Nepomechie,
``Open-chain transfer matrices for AdS/CFT,''
{\it JHEP} {\bf 0809}, 085 (2008)
[arXiv:0808.2629].

\bibitem{NR}
R. I. Nepomechie and E. Ragoucy,
``Analytical Bethe ansatz for the open AdS/CFT $SU(1|1)$ spin chain,''
{\it JHEP} {\bf 0812}, 025 (2008)
[arXiv:0810.5015].

\bibitem{Galleas}
W. Galleas,
``The Bethe ansatz equations for reflecting magnons,''
[arXiv:0902.1681].

\bibitem{Nep}
R. I. Nepomechie,
``Bethe ansatz equations for open spin chains from giant gravitons,''
{\it JHEP} {\bf 0905}, 100 (2009)
[arXiv:0903.1646].

\bibitem{CY2}
D.H. Correa and C.A.S. Young,
``Finite size corrections for open strings/open chains in planar AdS/CFT,''
[arXiv:0905.1700].

\bibitem{giants}
J. McGreevy, L. Susskind and N. Toumbas,
``Invasion of the giant gravitons from anti-de Sitter space,''
{\it JHEP} {\bf 0006}, 008 (2000)
[arXiv:hep-th/0003075];\\
M.T. Grisaru, R.C. Myers and O. Tafjord,
``SUSY and Goliath,''
{\it JHEP} {\bf 0008}, 040 (2000)
[arXiv:hep-th/0008015];\\
A. Hashimoto, S. Hirano and N. Itzhaki,
``Large branes in AdS and their field theory dual,''
{\it JHEP} {\bf 0008}, 051 (2000)
[arXiv:hep-th/0008016].

\bibitem{Ch} 
I.V. Cherednik, 
``Factorizing particles on a half line and root systems,''
{\it Theor. Math. Phys.} {\bf 61}, 977 (1984).

\bibitem{GZ}
S. Ghoshal and A.B. Zamolodchikov, 
``Boundary $S$-Matrix and Boundary State in Two-Dimensional 
Integrable Quantum Field Theory,''
{\it Int. J. Mod. Phys.} {\bf A9}, 3841 (1994) 
[arXiv:hep-th/9306002].

\bibitem{BK}
N. Beisert and P. Koroteev,
``Quantum Deformations of the One-Dimensional Hubbard Model,''
{\it J. Phys.} {\bf A41}, 255204 (2008).
[arXiv:0802.0777].

\bibitem{AB}
F.C. Alcaraz and R.Z. Bariev,
``Interpolation between Hubbard and supersymmetric $t-J$ models.
Two-parameter integrable models of correlated electrons,''
{\it J. Phys.} {\bf A32}, L483 (1999)
[arXiv:cond-mat/9908265].

\bibitem{EFGKK}
F.H.L. Essler, H. Frahm, F. G\"ohmann, A. Kl\"umper and V.E. Korepin,
{\it The One-Dimensional Hubbard Model} (Cambridge University Press, 
2005).

\bibitem{MuN}
R. Murgan and R.I. Nepomechie,
``$q$-deformed $su(2|2)$ boundary $S$-matrices via the ZF algebra,''
{\it JHEP} {\bf 0806}, 096 (2008)
[arXiv:0805.3142].

\bibitem{Sk}
E.K. Sklyanin, 
``Boundary conditions for integrable quantum systems,''
{\it J. Phys.} {\bf A21}, 2375 (1988).

\bibitem{AFr}
G. Arutyunov and S. Frolov,
``On string $S$-matrix, bound states and TBA,''
{\it JHEP} {\bf 0712}, 024 (2007)
[arXiv:0710.1568].

\bibitem{graded}
A. Foerster and M. Karowski,
``The supersymmetric t-J model with quantum group invariance,''
{\it Nucl. Phys.} {\bf B408}, 512 (1993);\\
A. Gonz\'alez-Ruiz,
``Integrable open-boundary conditions for the supersymmetric t-J 
model. The quantum group invariant case,''
{\it Nucl. Phys.} {\bf B424}, 468 (1994) 
[arXiv:hep-th/9401118];\\
R.H. Yue, H. Fan and B.Y. Hou,
``Exact diagonalization of the quantum supersymmetric $SU_q(n|m)$ 
model,''
{\it Nucl. Phys.} {\bf B462}, 167 (1996)
[{\tt cond-mat/9603022}];\\
M. Shiroishi and M. Wadati,
``Integrable Boundary Conditions for the One-Dimensional Hubbard 
Model,''
{\it J. Phys. Soc. Jpn.} {\bf 66}, 2288 (1997)
[arXiv:cond-mat/9708011];\\
X.-W. Guan, ``Algebraic Bethe ansatz for the one-dimensional Hubbard model with open boundaries,''
{\it J. Phys.} {\bf A33}, 5391 (2000)
[arXiv:cond-mat/9908054];\\
D. Arnaudon, J. Avan, N. Cramp\'e, A. Doikou, L. Frappat and E. 
Ragoucy,
``General boundary conditions for the sl(N) and $sl(M|N)$ open spin 
chains,''
{\it J. Stat. Mech.} {\bf P08005}, 1 (2004)
[{\tt math-ph/0406021}].

\bibitem{BGZZ}
A.J. Bracken, X.-Y. Ge, Y.-Z. Zhang and H.-Q. Zhou,
``Integrable open-boundary conditions for the $q$-deformed 
supersymmetric $U$ model of strongly correlated electrons,''
{\it Nucl. Phys.} {\bf B516}, 588 (1998) 
[arXiv:cond-mat/9710141].

\bibitem{MezN}
L. Mezincescu and R.I. Nepomechie, 
``Integrable open spin chains with nonsymmetric $R$ matrices,'' 
{\it J. Phys.} {\bf A24}, L17 (1991). 





\end{thebibliography}
\end{document}